\documentclass[journal]{IEEEtran}

\usepackage{graphicx}
\usepackage{placeins}
\usepackage{amsmath}
\usepackage[linesnumbered,ruled]{algorithm2e}
\usepackage{booktabs}

\ifCLASSINFOpdf
\else
\fi

\begin{document}
%
\title{Modeling Transient Negative Capacitance in Steep-Slope FeFETs}
%
%
%

\author{Borna~Obradovic,
        Titash~Rakshit,
				Ryan~Hatcher,
				Jorge~Kittl,
        and~Mark~S.~Rodder
\thanks{B. Obradovic, T. Rakshit, R. Hatcher, J. Kitttl and M. S. Rodder are with the 
Samsung Advanced Logic Lab, Austin TX (e-mail: b.obradovic@samsung.com)}

}
\maketitle

\begin{abstract}
We report on measurements and modeling of FE HfZrO/SiO2
Ferroelectric-Dielectric (FE-DE) FETs which indicate that
many of the phenomena attributed to Negative Capacitance can be
explained by a delayed response of ferroelectric domain
switching - referred to as Transient Negative Capacitance (TNC). 
No traversal of the stabilized negative capacitance
branch is required. Modeling is used to correlate the hysteretic
properties of the ferroelectric material to the measured
transient and subthreshold slope (SS) behavior. It is found that
steep SS can be understood as a transient phenomenon, present
when significant polarization changes occur. The experimental signature of
TNC is investigated, and guidelines for detecting it in measured data are outlined.
The technological implications of FE polarization switching are investigated, and
it is found that NCFETs relying on it are not suitable for high performance CMOS logic, due to 
voltage, frequency, and hysteresis limitations. Requirements for experimental evidence of stabilized S-curve behavior 
are summarized.

\end{abstract}

\begin{IEEEkeywords}
Ferroelectric, FeCap, Negative Capacitance, NCFET, FeFET
\end{IEEEkeywords}

%
\IEEEpeerreviewmaketitle

\section{Introduction}
\label{Intro}
\begin{figure}
\centering
\includegraphics[width=3.2in]{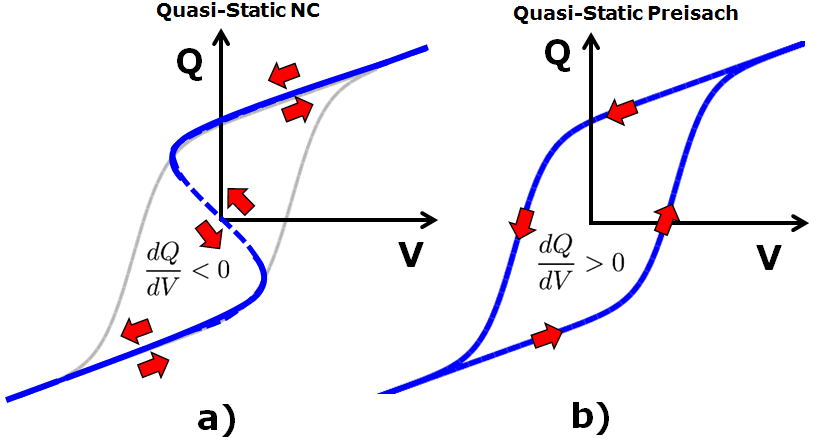}
\caption{The quasi-static Q-V curves for FeCaps are shown, based on the Preisach model (sub-plot b),
and the stabilized Negative Capacitance model (sub-plot a). Saturation loops are shown. In the Preisach
model, traversal of the saturation loop is always in the indicated direction; reversing the traversal direction
prior to full saturation creates minor loops. Quasi-static capacitance in the Preisach model  is always positive. 
In the stabilized NC model,
a non-hysteretic path is available. Traversal of the path is bi-directional. Quasi-static capacitance is negative
along the dashed portion of the path. }
\label{PreisachModel}
\end{figure}

%
%
%
%
\IEEEPARstart{N}{egative} capacitance has been postulated theoretically \cite{Datta} and extensively studied experimentally
(\cite{Sayeef}, \cite{ferro1}, \cite{SDattaEDL}).
The theoretical basis for Negative Capacitance (NC) is an assumed Energy-Charge (U-Q) relationship
with a region of negative curvature (U-Q ansatz) \cite{Datta}. As a consequence of the negative curvature,
a switching path with negative capacitance (Fig. \ref{PreisachModel}a) becomes available. This is in contrast to the
standard Preisach model of a ferroelectric capacitor (Fig. \ref{PreisachModel}b). However, this path
is unstable and not directly observable in stand-alone FeCaps. Arguments have been made 
\cite{Datta}  that
this unstable path can be stabilized by connecting the FeCap in series with a standard (positive U-Q curvature)
capacitor which satisfies specific matching criteria \cite{Stabilization}, 
thereby making the negative capacitance branch traversable. As a consequence, using such a capacitor
arrangement in the gate stack of a FET would result in increased stack capacitance and sub-60 mV/dec 
subthreshold slope (\cite{Datta}, \cite{Sayeef}). However, precisely how the stabilized state is established at the microscopic
level is unclear at the present, and challenging to explain. In the face of this difficulty, 
the goal of this paper is to investigate whether
some of the key experimental findings, including sub-60 mV/dec SS, can in fact be reproduced without the U-Q ansatz, consequently doing away
with the need for a microscopic model of the stabilized state. 
In this work, which is an extended version of \cite{Obradovic} (adding additional explanatory material), the alternative conceptual model proposed
is simply a Preisach Ferroelectric (\cite{Preisach}, \cite{Preisach2}) with slow ferroelectric (FE)
switching dynamics.
The presence of slow switching dynamics of ferroelectric domains is itself well known and characterized, with little ambiguity with regards to the microscopic model (\cite{ferro2}, \cite{ferro3}, \cite{Namlab}).

\FloatBarrier
\section{Modeling and Pulse-Based Verification}
\label{Modeling}

\begin{figure}
\centering
\includegraphics[width=3.0in]{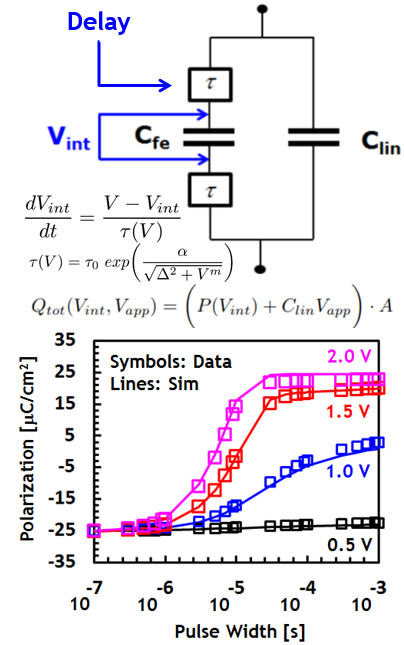}
\caption{The FeCap model is shown (top), with the associated model equations (center). The calibration
of the FE switching dynamics to measured data is shown at the bottom. An explicit delay model is used in conjunction with
a quasi-static turning point model for the FE polarization. A simple linear capacitor is used for the
non-FE polarization.}
\label{PVModel}
\end{figure}

The overall model for the FeCap consists of two components: a delayed ferroelectric polarization capacitance, and
a quasistatic non-ferroelectric capacitance. The ferroelectric response
is delayed due to the intrinsic switching dynamics of the Fe domains, while the non-ferroelectric component is
governed by fast electronic polarization. The ferroelectric response is modeled as a 
quasistatic Preisach FeCap combined with an explicit delay, using a dynamic turning-point model to describe the hysteresis (Fig. \ref{PVModel}, top). In this work, the delay is assumed to be due to the internal switching dynamics of 
ferroelectric domains, but no attempt is made to explain the detailed microscopic origin \cite{Namlab}.  The quasistatic parameters of the FeCap model are calibrated using the data of \cite{SDattaEDL},
as shown in Fig. \ref{PVModel}. In this work, $V_{int}$ (Fig. \ref{PVModel}) is governed by a first-order delay of the applied voltage across the FeCap, using a form of Merz's Law. (Fig. \ref{PVModel}, inset). The free charge on the capacitor plates is assumed to be quasi-static w.r.t. the total polarization (valid for continuum model and normal CMOS operation). 

\begin{figure}
\centering
\includegraphics[width=3.0in]{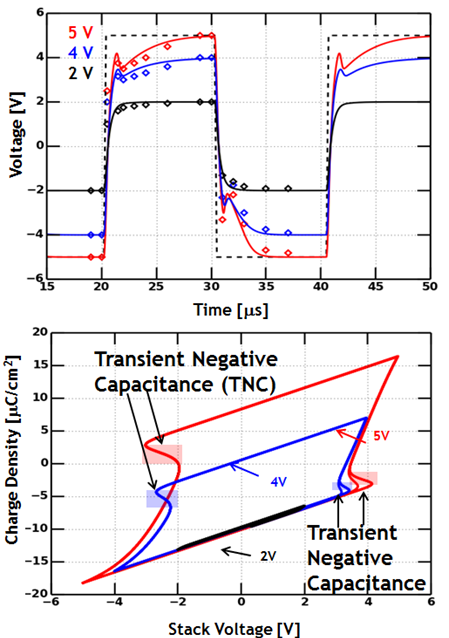}
\caption{The measured \cite{SDattaEDL} (symbols) and simulated (lines) transient responses for the ferroelectric capacitor stack
to bipolar square-pulses of various amplitudes are shown. The ``anomalous" spikes are diminishing in magnitude
with reducing input pulse amplitude (in both data and simulation). At a pulse voltage of 2V, the spikes are absent.
Varying spike magnitudes are observed as a function of the applied voltage pulses, for both positive
and negative pulses. The P-V trajectory is shown on the bottom, indicating regions of transient negative capacitance
while the stack voltage is decreasing.}
\label{Bipolar_VDD}
\end{figure}

The applicability of the model is tested by comparing predictions to pulse-train data of \cite{SDattaEDL}.
For bipolar switching, the square waveform ranges from -VDD to +VDD, exercising both the positive and
negative domain switching thresholds. Both the measured data and simulation (Fig. \ref{Bipolar_VDD}) exhibit an initial
voltage spike, followed by a more gradual rise to the peak pulse voltage. Arguing from the standpoint of a delayed ferroelectric
response, the ``spike" behavior can be understood as follows. The voltage across the stack initially rises quickly, with only the non-ferroelectric polarization responding. A $\mu$s or so later, the ferroelectric domains begin to switch, and the polarization increases dramatically. This is balanced by by an increase in current, which in turn drops more voltage across the access resistor, resulting in a brief drop in the voltage across the stack. The voltage then gradually rises as additional ferroelectric domains switch and the capacitors absorbs more charge.
The P-V trajectory (Fig. \ref{Bipolar_VDD}, right) shows regions of Transient Negative Capacitance (TNC) where
the voltage across the stack is decreasing.
Additional insight into the switching behavior is gained by examining the VDD behavior of the switching
in Fig. \ref{Bipolar_VDD}. It is apparent that the ``spike" effect is diminishing with decreasing pulse amplitude. While quite pronounced at 5V, it is completely absent at 2V (data and model). The reason is that at low pulse voltages the voltage drop across the FeCap is too small to trigger ferroelectric domains. From Fig. \ref{PVModel}, it is evident that very little ferroelectric switching happens in the voltage range of [-1.5V, 1.5V], which is the approximate voltage range across the FeCap during the 2V pulse. In the Verilog-A model, the delay is associated only with ferroelectric switching, so absent switching, there is negligible delay (and hence no spike). 

\begin{figure}[!ht]
\centering
\includegraphics[width=2.6in]{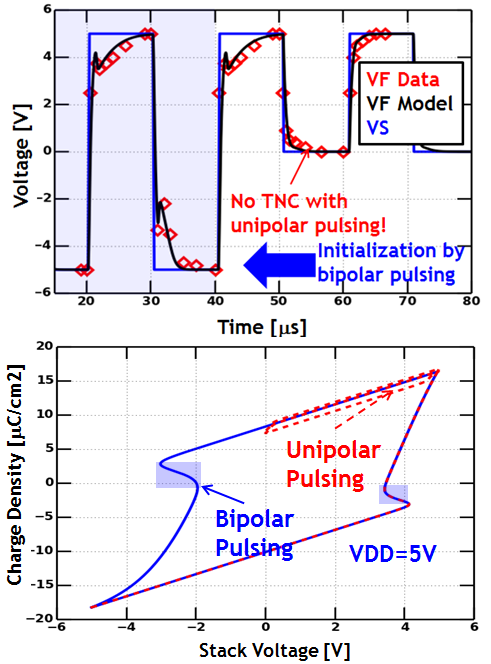}
\caption{The measured \cite{SDattaEDL} and simulated transient response for the ferroelectric capacitor stack
to unipolar square-wave pulses is shown (left). The stack is initialized by a sequence of bipolar pulses (shaded blue region),
followed by a sequence of unipolar pulses. On the first unipolar pulse, the ``anomalous" spike is observed. On subsequent
pulses, the spike is absent. The P-V trajectory is shown on the bottom; unipolar pulsing is seen to produce a tight minor loop with negligible $\Delta P$. }
\label{Unipolar}
\end{figure}

For unipolar switching, the square input waveform is modified to range from 0 to VDD. The results are qualitatively different than
for the bipolar case. As can be seen in Fig. \ref{Unipolar}, the ``spike" is observed only on the first unipolar pulse of the waveform.
Subsequent pulses exhibit no spike at any voltage. This behavior
is reproduced in simulation. The reason for the behavior
of the stack voltage is apparent from Fig. \ref{Unipolar} (bottom) . While the polarization undergoes large changes during bipolar
switching, the only large polarization change during unipolar switching is during the transition from bipolar to unipolar mode.
After this initial pulse, the FeCap operates on a tight minor loop, and the polarization changes are small and mostly
due to non-ferroelectric polarization.
As previously discussed, non-ferroelectric polarization is essentially quasistatic, so there is a negligible amount of delayed polarization. Hence, no spike is produced.

\FloatBarrier
\section{Impact on Subthreshold Slope}

The technological impetus for investigating the NC-effect is the potential improvement 
in FET subthreshold slope (SS). As argued in \cite{Datta}, a negative capacitance gate
layer can result in a sub-60 mV/dec SS, by introducing ``amplification" into the 
surface potential ($\psi$). Specifically, if the applied gate bias is V${_g}$, the 
long channel slope of the I$_d$-V$_g$ curve is proportional to $d\psi/d V_g$. With ordinary capacitors, this
derivative is never greater than unity (equal to unity only for long-channel, fully-depleted
devices). As shown in \cite{Datta}, NC effects can push the derivative beyond unity,
thereby reducing the SS below the usual theoretical limit.

\subsection{Conceptual Model}
\label{sec:conceptual_model}

\begin{figure}[!ht]
\centering
\includegraphics[width=3.6in]{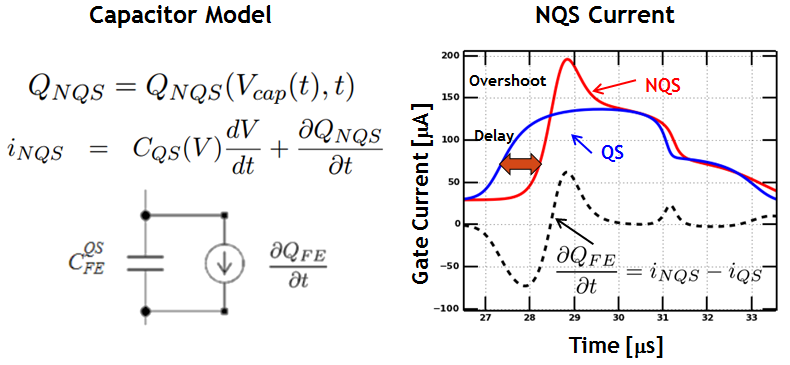}
\caption{The NQS capacitance model (left) and the current through an FeCap (right)
during a rising voltage ramp are illustrated. The QS
response is obtained by assuming fast internal dynamics, while the NQS response uses 
realistic (and slow) dynamics. The difference of the true NQS response and the QS response
yields the $i_{NQS}$ current used by the circuit model (left) }
\label{NQS_current}
\end{figure}

In order to gain insight into the effect of the non Quasi-Static TNC behavior of FeCaps, a simple
conceptual model is considered first. The charge on the FeCap is a non Quasi-Static (NQS) 
function of voltage and time,
having an explicit temporal dependence due to internal polarization dynamics. In terms of a circuit
element description, the NQS FeCap must therefore be modeled as shown in Fig. \ref{NQS_current} (left).
The form of the NQS current component in the circuit model of Fig. \ref{NQS_current} is governed
by the polarization dynamics of the FeCap and depends on the details of the model used. Using the
previously described model, the current response to a voltage
ramp is shown in Fig. \ref{NQS_current} (right).

\begin{figure}[!ht]
\centering
\includegraphics[width=2.8in]{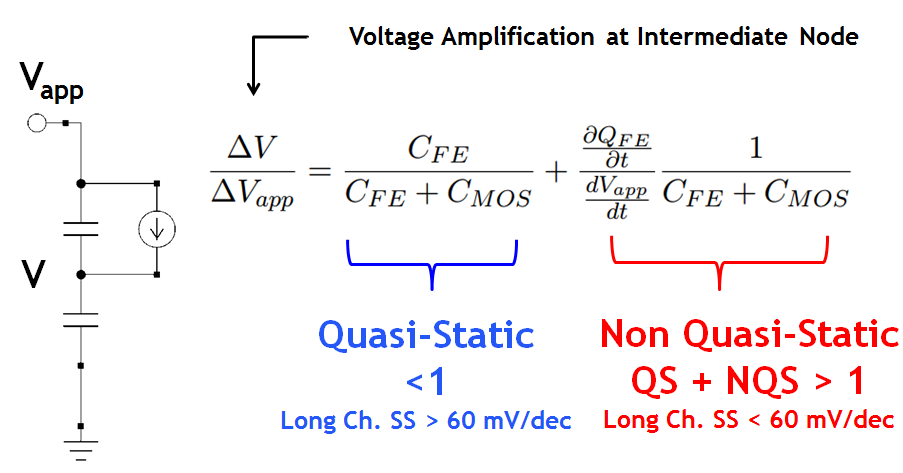}
\caption{The amplification at the intermediate node of a capacitive divider with an NQS
cap element is illustrated. The amplification term consists of two parts: a QS term, which
is always less than unity, and an additional NQS term, which may push the sum of the QS and NQS
amplification terms beyond unity. }
\label{amplification_concept}
\end{figure}

The application of the circuit model for the FeCap of Fig. \ref{NQS_current} to a simple capacitive
voltage divider is shown in Fig. \ref{amplification_concept}. The voltage at the intermediate node in
Fig. \ref{amplification_concept} is then
obtained as:
\begin{eqnarray}
C_{MOS} \frac{dV}{dt} + C_{FE} \frac{d}{dt} (V-V_{app}) - \frac{\partial Q_{FE}}{\partial t} = 0 \\
\frac{dV}{dt} = \frac{C_{FE}}{C_{FE} + C_{MOS}} \frac{dV_{app}}{dt}
+ \frac{\partial Q_{FE}}{\partial t} \frac{1}{C_{FE}+C_{MOS}} \\
\end{eqnarray}

It should be noted that all the $C_{FE}$ and $C_{MOS}$ terms in the above equations are functions of the
instantaneous voltages across the corresponding capacitors (but not explicitly functions of time; this is handled
by the NQS current term $\frac{\partial Q_{FE}}{\partial t}$), and are not considered to be constant. 

\noindent Dividing through by the derivative terms and approximating increments using $\Delta V \approx \frac{dV}{dt} \Delta t$,
one obtains:
\begin{equation}
\frac{\Delta V}{\Delta V_{app}} = \frac{C_{FE}}{C_{FE} + C_{MOS}}
+ \frac{\frac{\partial Q_{FE}}{\partial t}}{\frac{dV_{app}}{dt}} \frac{1}{C_{FE}+C_{MOS}}
\label{amplification}
\end{equation}

\noindent Since Eqn. \ref{amplification} relates the incremental change of the intermediate node potential w.r.t. an incremental
change of the applied voltage, it is the sought-after amplification. Under quasi-static conditions (i.e. all NQS currents are very small because the internal dynamics of the FE cap can track a slow voltage ramp) the 2nd term
in Eqn. \ref{amplification} vanishes (i.e. $ \frac{\partial Q_{FE}}{\partial t} \rightarrow 0$). In this case, 
the amplification is always less than unity, approaching unity in the limit of $C_{FE} >> C_{MOS}$. This is
the ``normal'' MOS regime, in which SS $>$ 60 mV/dec. It should be noted that this result holds for arbitrary non-linear capacitive
behavior of $C_{FE}$ and $C_{MOS}$; the only requirement for ``normal'' MOS behavior is the quasi-static condition.
With non quasi-static conditions, however, the 2nd term must be taken into account, as highlighted in 
Fig. \ref{amplification_concept} . For cases where $ \frac{\partial Q_{FE}}{\partial t}$ is positive (and assuming that $\frac{dV_{app}}{dt}$ is positive), amplification is increased.
It should also be noted that the mere presence of the NQS term in Eqn. \ref{amplification} and 
Fig. \ref{amplification_concept} does not guarantee that amplification exceeds unity.
Only if the 2nd term is sufficiently large and positive can amplification exceed unity. 
If this is the case, SS may be less than 60 mV per decade. The key is the behavior of the NQS current, given by $ \frac{\partial Q_{FE}}{\partial t}$.
Clearly, extreme simplification assumptions were made in obtaining this result w.r.t. the MOS model: it was treated
as a single, lumped capacitance. In reality, it is the surface potential hidden inside this lumped model that is
of interest, not just the potential of the intermediate node of Fig. \ref{amplification_concept}. 
Two and three-dimensional effects were ignored.
The details
of the FET model are discussed next.

\subsection{FET Simulation}
\label{sec:fefet_sim}
\begin{figure}[!ht]
\centering
\includegraphics[width=1.3in]{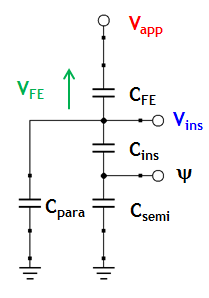}
\caption{The voltage division across an FET gate stack is illustrated. Of key interest is the presence
of amplification at the surface potential node $\psi$. }
\label{divider_FET}
\end{figure}

While the conceptual model is helpful in arguing for the possibility
of amplification and sub-60 mV/dec FET behavior, a more detailed accounting of the FET behavior requires
simulation with an appropriate FeCap and FET model. The capacitive divider representing the full stack
of the FET is illustrated in Fig. \ref{divider_FET}.

\begin{figure}[!ht]
\centering
\includegraphics[width=3.6in]{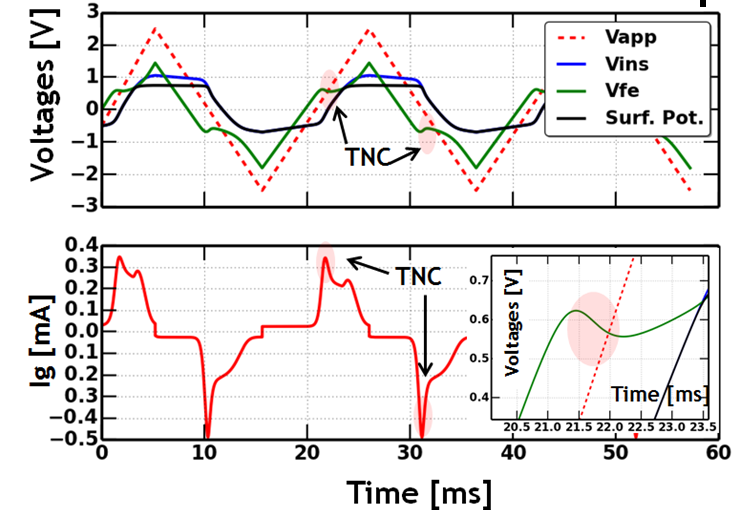}
\caption{The simulated waveforms of an FET driven by a triangular voltage wave are shown. The
curve colors correspond to the potentials at nodes of Fig. \ref{divider_FET}. Amplification is seen
periodically at brief intervals, shaded with red ovals. It can also be seen that the regions
of amplification are also regions of TNC.}
\label{transient_FET}
\end{figure}

The behavior of the full FET with an applied triangular-wave voltage waveform is simulated using the
FeCap model of Sec. \ref{Modeling} with HSPICE. The resulting waveforms are illustrated in 
Fig. \ref{transient_FET}. Regions of TNC are clearly evident: the voltage across the FeCap (green curve)
exhibits brief ``counter-trend'' behavior, in which the voltage drops (rises) during periods when 
the applied gate voltage rises (drops). This is the same behavior as previously 
noted in Fig. \ref{Bipolar_VDD}. The reduction in magnitude of the FeCap voltage implies that the potential
at the linear capacitor node $V_{ins}$ must be increasing faster than the applied voltage. Viewed from
the context of the conceptual model of Sec.\ref{sec:conceptual_model}, there is a burst of current arising
from the delayed FE polarization which rapidly charges the capacitors below the FeCap, 
increasing the potential at all floating nodes of Fig. \ref{divider_FET}.
This potential increase results in a reduction of voltage across the FeCap while its charge is increasing, 
i.e. TNC. Thus, in this view,
TNC is seen as a side effect of amplification, not its cause.  

\begin{figure}[!ht]
\centering
\includegraphics[width=3.0in]{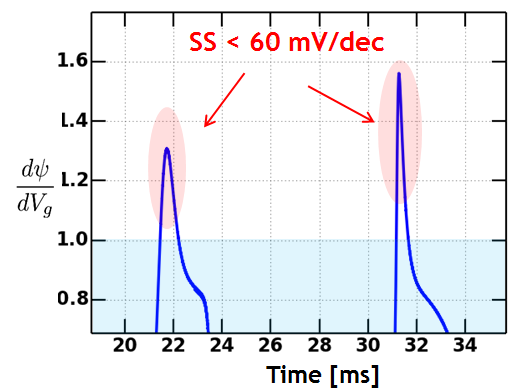}
\caption{Voltage amplification at the surface potential node $\psi$ is shown, expressed as
$\frac{d\psi}{dVg}$. The values are obtained by numerical differentiation of data from Fig. \ref{divider_FET}.
Periodic regions of strong amplification are observed (only one period shown). }
\label{amplification_FET}
\end{figure}

The values for amplification $\frac{d\psi}{dVg}$ can be obtained directly from Fig. \ref{divider_FET}, and
are illustrated in Fig. \ref{amplification_FET}. Regions where $\frac{d\psi}{dVg} > 1$ may correspond to
regions where $SS < 60$ mV/dec for long-channel FETs. 

\begin{figure}[!ht]
\centering
\includegraphics[width=2.5in]{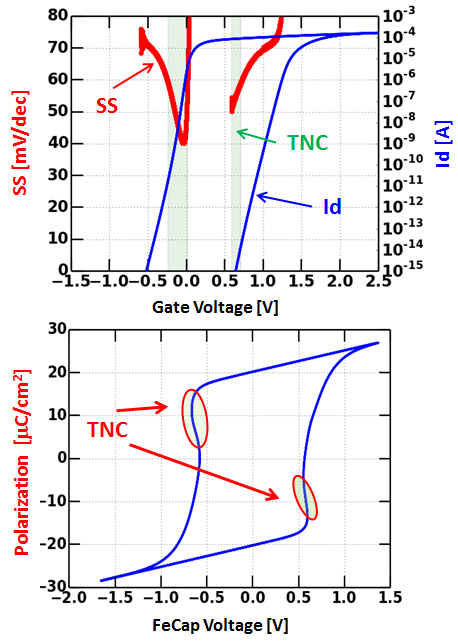}
\caption{The Id-Vg of the FET (top) and the corresponding P-V trajectory of the FeCap are shown
for the input waveform of Fig. \ref{divider_FET}. Strong hysteresis is observed in Id-Vg, and
the SS is indeed below 60 mV/dec in both the forward and reverse sweep. 
Correspondingly, the P-V trajectory
shows regions of TNC.
Due to the placement of the FET Vt, however,
the sub-60 region of the forward sweep takes place at very low current levels. In a real FET (with a BTBT leakage floor),
sub-60 mV/dec SS would not have been observed in the forward direction.}
\label{sub60_IdVg}
\end{figure}

The reason that  $\frac{d\psi}{dVg} > 1$ does not 
guarantee sub-60 SS long-channel behavior is due to the fact that regions of  $\frac{d\psi}{dVg} > 1$ may
not be aligned with sub-threshold values of the surface potential $\psi$, i.e. the $V_t$ of the FET is 
not properly selected. This is in fact seen in the current example, as shown in Fig. \ref{sub60_IdVg}.

\subsection{Hysteresis}

\begin{figure}[!ht]
\centering
\includegraphics[width=2.65in]{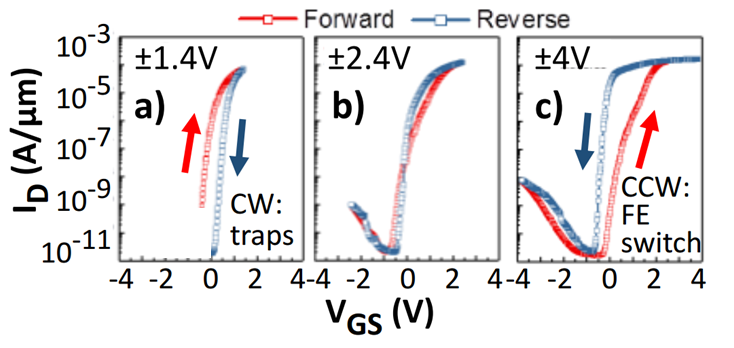}
\caption{The behavior of trap-induced (CW) and FE switching-induced (CCW) hysteresis
is illustrated. At low peak Vg values (leftmost figure), no FE switching takes place and
the hysteresis has a trap-like character. Conversely, at high peak Vg (rightmost figure),
strong FE switching takes place, and the hysteresis has FE-like character. At intermediate
voltages, there is a partial cancellation of the two hystereses, i.e. the Vt-shift induced
by each is partially compensated. At a precisely chosen peak Vg, hysteresis appears to be
fully canceled (though obviously only approximately so). }
\label{trap_hysteresis}
\end{figure}

As seen in Fig. \ref{sub60_IdVg}, sub-60 mV/dec SS is accompanied by significant Id-Vg hysteresis.
This is an inescapable result in the presence of polarization switching, resulting in an effective 
Vt-shift of the underlying FET. Virtually all published measurements of sub-60 mV/dec FETs show
this strong correlation of SS improvement and Id-Vg hysteresis.
Some reported works do indeed report
hysteresis-free sub-60 mV/dec behavior, but a closer inspection reveals the
behavior to be only approximately hysteresis-free. Measurements indicate that this is caused
by a balance of trap-induced, clockwise hysteresis, and FE switching-induced counter-clockwise
hysteresis. This effect is illustrated in Fig. \ref{trap_hysteresis}.

\subsection{Frequency and Peak Voltage Dependence}

\begin{figure}[!ht]
\centering
\includegraphics[width=2.5in]{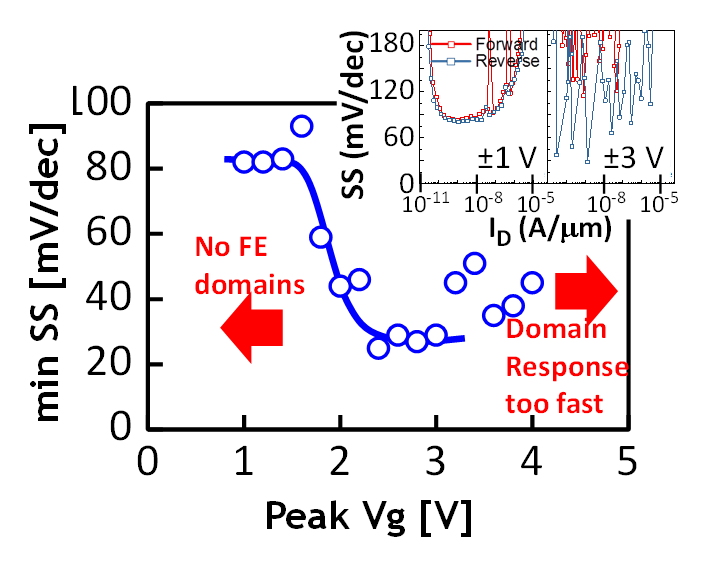}
\caption{The min SS obtained for an NCFET as a function of the peak Vg is shown. At low voltages, the SS
is seen to be above 60 mV/dec, and no NC-related enhancement is observed. As the peak voltage is increased,
the min SS is reduced below 60 mV/dec, clearly showing NC-like behavior. Further increases in peak Vg result
in increased SS however, as the response time of the FE polarization becomes significantly faster than
the applied voltage waveforms. The measured SS vs. Id is shown in the inset.}
\label{peak_voltage}
\end{figure}

The conceptual model of Sec. \ref{sec:conceptual_model} suggests that the magnitude of voltage amplification
depends on the NQS current. Since the proposed NQS current mechanism is simply a delayed polarization current, it is
clear that SS should depend on the amount of switched polarization. Correspondingly, the SS behavior of the NCFET
depends on the peak switching voltage (since the latter controls the amount of switched polarization $\Delta P$),
but also on the rate at which the applied voltage across the NCFET changes, since the polarization dynamics
can only keep up with applied voltage ramps that are not much faster than the polarization delay time.
The dependence of the min SS of a single NCFET across a range of peak voltages is shown in Fig. \ref{peak_voltage}.

The measured min SS vs. peak Vg follows a ``U-shaped'' curve, due to two competing effects. At low peak
voltages, the voltage across the FeCap stays below the coercive voltage V$_c$, thereby limiting FE domain
switching, resulting in minimal (or no) $\Delta P$, and a correspondingly negligible $\frac{\partial Q_{FE}}
{\partial t}$ of Eqn. \ref{amplification}. Under such circumstances, SS is expected to be $>$ 60 mV/dec, 
and indeed measured data confirms this. As the peak voltage is increased, the min SS is seen to dip below
60 mV/dec, to as low as 30 mV/dec. In this regime, the voltage across the FeCap does exceed V$_c$, and
large values of $\Delta P$ and $\frac{\partial Q_{FE}}{\partial t}$ are possible. Finally, further increases
in the peak voltage show a gradually increasing min SS. In this regime, there is little or no further
increase in the switched FE polarization (peak voltage in the FeCap far exceeds V$_c$). However, the higher
field in the FeCap does result in faster domain switching. This increased switching rate with a fixed
input waveform at the gate results in nearly Quasi-Static behavior of the FeCap, reducing the current
overshoot in Fig. \ref{NQS_current} and the corresponding NQS term in Eqn. \ref{amplification}.

\begin{figure}[!ht]
\centering
\includegraphics[width=2.5in]{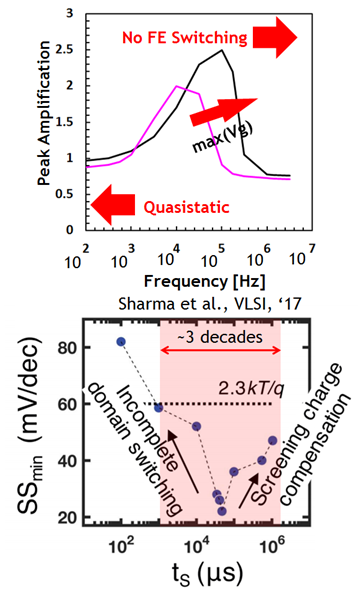}
\caption{The simulated (top) and measured (bottom) dependence of amplification (SS) on the
measurement rate is illustrated. The top figure illustrates the long-channel 
surface potential amplification as a function of the frequency of the input triangular wave,
across a range of peak voltages. In this context, amplification is defined as the ratio
of $\frac{d\psi}{dV_g}$ at frequency f to the quasi-static $\frac{d\psi}{dV_g}$. 
Strong amplification is seen in a relatively narrow band around the
peak, spanning about two decades in frequency. The bottom figure shows the measured
min SS as a function of the measurement step time. A peak (trough) is observed in the min SS,
with a witdth of about three decades.}
\label{freq_dep}
\end{figure}

In general, it should be expected that the sub-60 SS behavior (or conversely, surface potential
amplification) should be strongly frequency dependent, as illustrated in Fig. \ref{freq_dep}.
The top plot of Fig. \ref{freq_dep} shows the simulated amplification factor for a single NCFET,
as a function of the frequency of the applied triangular waveform. There is a clear peak, with
the amplification decaying at high frequencies (where domain switching is too slow), but also at
very low frequencies, where the ferroelectric behavior is quasi-static (as discussed in the context of the
peak-Vg dependence). The position of the peak is V$_g$-dependent, due to the V$_g$ dependence of the
FE domain switching speed. The simulated width of the amplification peak spans about two decades in frequency. 
This behavior is qualitatively reproduced in measured data, as shown in the
bottom plot of Fig. \ref{freq_dep}. The measured min SS is seen to be strongly frequency-dependent
(the x-axis parameter is the voltage step time during the triangular ramp, serving here as a proxy for the
inverse of the sweep frequency). It should be noted that the bandwidth of the measured sub-60 response
is somewhat wider than that of the simulation, spanning about three decades (as compared to about
two for simulation). This suggests that the modeled polarization dynamics are not capturing the full
complexity of the FE behavior. One obvious modeling simplification is the notion that all the FE
domains follow the same switching dynamics. There is no $a priori$ reason for this, and it is not unreasonable
to expect that domains with different threshold voltages will likewise have different switching dynamics.
This is in fact evident in the measurements of \cite{Namlab}, which indicate that there is roughly one
order of magnitude spread in switching times of various domains at the same applied voltage. This is one
of several indications that treating the FeCap as a continuum is an oversimplification, and that explicitly
accounting for domain discreteness is necessary.

\subsection{Effect of Domain Discreteness}

As mentioned in the previous section, treating the FE layer as a continuum is a great model simplification.
In fact, HfZrO (and PZT) layers are granular, with a relatively small number of discrete ferroelectric domains.
Polarization of the FE layer cannot change in a continuous fashion; it naturally occurs only in discrete jumps.
Thus, it should not be surprising that measurements of SS frequently show ``noisy'' behavior that appears to
be consistent with discrete domain switching. An example is shown in the inset of Fig. \ref{peak_voltage}, in
which it can be seen that SS has a large random component under conditions in which sub-60 behavior is observed.
This random component is not easily dismissed as measurement noise of low-level currents; the same current levels
give rise to nearly noise-free SS under conditions in which SS $>$ 60 mV/dec (Fig. \ref{peak_voltage}, inset:
low peak-V$_g$ values in which no FE switching occurs have SS $>$ 60 mV/dec and small noise; high peak-V$_g$
measurements result in sub-60 mV/dec, accompanied by significant SS noise). 
Additionally, the relatively large average grain size suggests that only a small number of FE domains can be
present on a scaled FET. A typical finFET might have a gate area of approximately 3000 $nm^2$ (two fins);
this would suggest an expectation value of fewer than 10-20 FE domains per FET. Not only is the number of 
FE domains expected to vary from FET to FET, but their size and threshold voltages as well. Thus, examining
a statistical, multi-domain FeCap model is necessary.
Following the Preisach \cite{Preisach} approach of treating FeCaps as collections of independent FE domains,
a multi-domain can be simply constructed as a parallel combination of a number of individual ``hysterons,'' 
where the hysterons are represented by the ideal box-like P-V hysteresis relations.
While the full impact of statistical variations is a subject worthy of further investigation,
a case of specific interest is illustrated herein. The grain size distribution of \cite{Namlab} suggests
that FeCaps consisting of a large domain and a collection of small domains are not unlikely. It is of particular interest then to see the electrical impact of switching a large domain. This is modeled (for example) by an FeCap in which
25\% of the area is occupied by a single domain (a hysteron), while the remaining 75\% consists of a large
number of small domains. The collection of small domains is treated as comprising a continuum, rather than
consisting of individual domains (this is done to highlight the effect of the single large domain).  

\begin{figure}[!ht]
\centering
\includegraphics[width=3.2in]{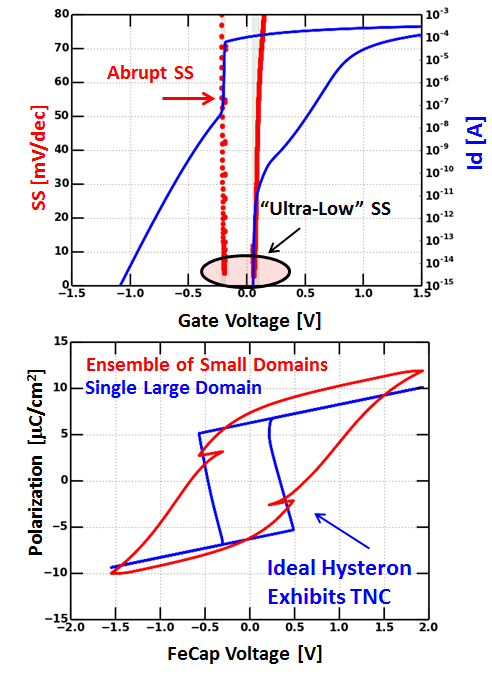}
\caption{The Id-Vg curves (top) and P-V trajectories (bottom) for the FET with a 25-75 FE
layer (25\% of area is single large domain, 75\% consists of many small domains) are shown.
The presence of the large domain induces a very sharp jump in current, resulting in a near-zero
SS. The P-V trajectory indicates that the TNC effect is driven entirely by the large domain. }
\label{spike}
\end{figure}

It is clear from Fig. \ref{spike} that the presence of the large domain has a significant impact
on the Id-Vg characteristics of the FET. Specifically, switching of the large domain produces a
very strong TNC effect, resulting in a near-abrupt change in channel current (SS $<$ 5 mV/dec).
Had this large domain be subdivided into many smaller domains, the resulting current overshoot 
would have been distributed over many events, producing a more continuous, albeit much less dramatic
drop in SS. Numerical experiments suggest that SS values much below 30 mV/dec are not achievable 
using the continuous model: large discrete domains appear to be necessary.
The highly abrupt nature of the SS improvement seen in Fig. \ref{spike} is commonly observed
in measurements. In fact, most measurements show discrete jumps in SS, rather than continuous sub-60
behavior, but only some demonstrate exceedingly low SS values, such as those of Fig. \ref{spike}.
It is suggested here that these measurements correspond to FETs with a small number of large FE domains.

\FloatBarrier
\section{Distinguishing TNC from Stabilized S-Curve}

Given the dramatic differences in CMOS applicability of FETs operating on a stabilized S-curve
vs. those operating by FE switching, it is important to be able to clearly identify
which type of sub-60 SS behavior is manifesting in a given experiment. A simple set
of distinguishing conditions is shown in Table \ref{tab:distinguish}.

\begin{table}[htbp]
  \centering
  \caption{TNC vs. S-Curve Distinguishing Characteristics}
    \begin{tabular}{cccc}
    \toprule
    \textbf{Conditions} & \textbf{TNC} & \textbf{S-Curve} & \textbf{Experiment To Date} \\
    \midrule
    Peak V$_{FE}$ $<<$ V$_C$ & No & Yes & No \\
    Peak V$_{FE}$ $>$ V$_C$ & Yes & Yes/No & Yes \\
    f $<$ f$_{FE}$ & Yes & Yes & Yes \\
    f $>>$ f$_{FE}$ & No & Yes & No \\
    Discrete Jumps & Yes & No & Yes \\
    Hysteresis & Yes & No & Yes \\
    \bottomrule
    \end{tabular}%
  \label{tab:distinguish}%
\end{table}%

In Table \ref{tab:distinguish}, V$_{FE}$ is the peak voltage across the FE capacitor, V$_C$
is the coercive voltage, f is the sweep frequency, and f$_{FE}$ is the response frequency
of the FE layer. In order to unequivocally demonstrate S-curve behavior, the FET should demonstrate
sub-60 mV/dec SS in the absence of FE switching. As shown in Table \ref{tab:distinguish}, this requires
sub-60 operation at low applied peak voltages (Peak V$_{FE}$ $<<$ V$_C$), at sufficiently high
switching frequencies (f $>>$ f$_{FE}$), and with no hysteresis. The voltage and hysteresis criteria
are relatively straightforward to apply in measurements; the value of V$_C$ can be obtained from the
P-V characteristics of the FE layer, and the peak voltage can be kept well below that value.

\FloatBarrier
\section{Conclusion and Future Direction}

A theoretical model of NCFETs which explains sub-60 mV/dec SS as a consequence of slow FE polarization
switching has been proposed. The predictions of the model are in qualitative or even quantitative agreement
with several classes of measured data. The measured data includes transient pulses across FE-DE stacks for
various voltages and polarities, as well as measurements of NCFET SS across a wide range of voltage and sweep
rate conditions.
The key feature of the model is that the relatively slow FE domain switching dynamics result in NQS
behavior of the FE layer. It was shown analytically that gate stacks with NQS components can exhibit voltage amplification,
or equivalently, improved SS. The delay in the FE polarization response relative to the applied gate
voltage produces NQS bursts of gate current, which can result in sub-60 SS. No explicit negative capacitance
is required for this enhancement to take place. In fact, the quasi-static FE capacitance is shown to always
be positive, even though voltage amplification in the gate stack implies a transient form of negative capacitance.
It was
 also argued that the transient nature of this enhancement results in significant limitations w.r.t. the
application of NCFETs to CMOS logic. Specifically, the FE polarization response time of HZO-based dielectric films
is limited to the order of $\mu$s or slower, rendering sub-60 SS behavior at CMOS circuit speeds impossible.
Furthermore, large applied voltages are generally required to produces sub-60 SS behavior; in order for FE domains
to switch, the voltage across the FE layer must be comparable to the coercive voltage. For thin films (5nm or less),
the coercive voltages are approximately 0.5V. Given the voltage division between the FE and DE layers, applied gate
voltages significantly in excess of 1V are required for sub-60 SS, even for thin-film NCFETs.

It was furthermore shown that the discrete domain nature of FE films is responsible for the nearly abrupt
jumps in drain current observed in some measured data. Individual domain switching events of large area domains result in large and sudden changes in polarization, leading to nearly abrupt changes in the drain current. 
NCFETs which contain a small number of large domains can manifest sub-60 SS even at very low sweep rates
(i.e. nearly quasi-static), but still not at frequencies exceeded the response frequency of domain switching.

While the NQS model of NCFETs presented in this paper is in good agreement with observed FET behavior to date, this
does not necessarily imply that sub-60 mV/dec SS is possible only by relying on delayed polarization switching. The
desired NCFET based on the stabilized S-curve may be independently achievable.
However, it is important to distinguish polarization switching NCFETs from stabilized S-curve NCFETs when analyzing measured data. For this purpose, a set of
guidelines has been presented, suggesting how to distinguish transient NC from the desired stabilized S-curve.
Specifically, in order to be confident in the S-curve interpretation of a measured sub-60 mV/dec SS, the
experiment must prevent any possibility of FE domain switching. This includes operating at low voltages 
(well below the coercive voltage of the FE layer) and at frequencies much higher than the response frequency
of the FE domains. If sub-60 mV/dec SS is indeed observed under conditions which preclude FE switching, 
it will be possible to claim that stabilized S-curve operation has in fact been achieved.



%



\ifCLASSOPTIONcaptionsoff
  \newpage
\fi



%
\FloatBarrier

%


\begin{IEEEbiographynophoto}{Borna Obradovic}
Borna Obradovic:
Borna Obradovic received the Ph.D. degree in Electrical Engineering from the University of Texas at Austin in 1999. He has worked in various aspects of microelectronics CAD at Intel, Texas Instruments, and most recently the Samsung Advanced Logic Lab.  His most recent work involves all aspects of hardware for neuromorphic computing, from process and device design to system simulation and software development.
\end{IEEEbiographynophoto}

\begin{IEEEbiographynophoto}{Titash Rakshit}
Titash Rakshit is a Principal Engineer at Samsung’s Advanced Logic Lab. He received his PhD degree in Electrical and Computer Engg from Purdue University in 2004. His thesis involved predicting negative differential resistance at the semiconductor-molecule interfaces. After graduation, he worked at Intel Corp. where he was part of the research team that developed the industry first demonstration scaled finFET technology. Since 2014, he has been working at Samsung Advanced Logic Lab focusing on future technology and systems roadmap. 
\end{IEEEbiographynophoto}

\begin{IEEEbiographynophoto}{Ryan Hatcher} 
Ryan Hatcher received B.A. and M.S. degrees in physics from Wake Forest University and a M.S. degree in computer science and Ph.D. in physics from Vanderbilt University.  In 2007, he joined Lockheed Martin’s Advanced Technology Laboratories where he drove a broad portfolio of research programs in advanced sensors and electronics. In 2013, he joined Samsung’s Advanced Logic Laboratory where he has focused on advanced logic technologies including both conventional CMOS as well as more exotic beyond-CMOS options. 
\end{IEEEbiographynophoto}

\begin{IEEEbiographynophoto}{Jorge A. Kittl}
Jorge A. Kittl is currently Vice President at the Advanced Logic Lab, Samsung, Austin, Texas, and a part-time Professor at the Dept. of Physics, KU Leuven, Belgium. He was Chief Scientist at imec between 2007 and 2012. Before that he was with Texas Instruments from 1993 to 2007 where he held several positions in different areas of R\&D, manufacturing and business management. He received his M. Sc. (‘87) and Ph.D. (‘91) degrees in Applied Physics from Caltech, and was a post-doctoral researcher at Harvard University. He has authored or co-authored over 200 publications.
\end{IEEEbiographynophoto}

\begin{IEEEbiographynophoto}{Mark S. Rodder}
Mark Rodder received his Ph.D. degree from MIT in 1987. He joined Texas Instruments where he contributed to device design/integration for many CMOS generations; he was promoted to TI Fellow in 1998. He is now Senior VP at Samsung Semiconductor, Inc., heading the Advanced Logic Lab in Austin, TX. He has 100 granted patents, has served on several conference committees including serving as the IEDM Short-Course Chair, and is an IEEE Fellow.
\end{IEEEbiographynophoto}




\end{document}